# Lithium intercalation drives remarkable mechanical properties deterioration in bulk and single-layered phosphorus: A first-principles study

Gao Xu,[a] Yanyu Liu,[b] Jiawang Hong,[b,*] and Daining Fang[a,c]

[a]*State Key Laboratory for Turbulence and Complex Systems, College of Engineering, Peking University, Beijing 100871, PR China*

[b]*School of Aerospace Engineering, Beijing Institute of Technology, Beijing 100081, China*

[c]*Institute of Advanced Structure Technology, Beijing Institute of Technology, Beijing 100081, PR China*

[*]Corresponding author

*E-mail address*: hongjw@bit.edu.cn

**Abstract:**

It is of critical importance to understand the mechanical properties change of electrode materials during lithium intercalation in the mechanical design of Li-ion batteries, for the purpose of the high reliability and safety in their applications. Here, we investigated the mechanical properties of both bulk and single layer phosphorus during the lithium intercalation process by using the first-principles calculations. Our results show that the Young's modulus of bulk and layered phosphorus strongly depends on the lithium intercalation. The mechanical bearing capacities, such as critical strain and stress, are significantly reduced by several times after lithium intercalation in both bulk and single layer phosphorus, which may reduce the reliability of Li-ion batteries. Our findings suggest that this remarkable mechanical properties deterioration during Li intercalation should be considered carefully in the mechanical design of Li-ion batteries, in order to keep they working reliably and safely in the charge-discharge process.

**Introduction**

With the rapid development of consumer electronics such as mobile devices and electric vehicles, rechargeable battery with high capacity and rate capacity has attracted tremendous attentions. Li-ion batteries (LIBs), one of the most widely studied rechargeable batteries, is a critical enabler for the next generation energy technology due to its advantages in portability and energy efficiency [1-3]. The growing interest in LIBs has greatly accelerated the development of electrode materials. Recently, most investigations on electrode materials are focused on the electrical and electrochemical properties. [4-7] However, the mechanical properties of electrode materials, such as elastic modulus and mechanical failure behaviors, [8-11] are less investigated. As is known, in LIBs, the volume change of the electrode active material during charging and discharging is one of the most critical factors for the stability of batteries. [12,13] Repeated expansion and contraction of the various phases due to Li diffusion in and out of the negative electrode during charge and discharge cycles can ultimately result in particle fracture and structural damage of electrode particles. This results in fragmentation of electrode particles, which is an important cause of capacity fading of batteries. [14-16] In addition, the other mechanical properties, such as the critical strain of failure, maximum permissible or mechanical strength deformation of electrode materials, are also important for its application. [17-20]

Mathematical models of deformation and corresponding stress fields during Li diffusion in and out of idealized electrode particle geometries have recently been developed. [21-24] These studies provided useful models for understanding structural changes in Li-ion batteries due to Li diffusion. In all of these continuum-level models, the intrinsic mechanical properties of electrodes, such as Young's modulus and Poison's ratio are assumed constants, independent of Li concentration. Actually, the mechanical properties of electrodes changes in different Li concentration, such as in general anode material Si [25] and graphite. [26]Yang et al. considered this effect and proposed the linear relationship between Young's modulus and the solute concentration. [27]This linear relationship was also shown in recent first-principles study, such as Li-Si phase [25] and alloy negative electrodes for Na-ion batteries [28], in which the relation of Young's modulus and Li/Na concentration approximate obeys the linear model. In addition to Young's modulus, the critical failure stress or critical failure strain were also deemed to be constant during lithium concentration change in mechanical models and LIBs designs. [29-31] However, this could bring serious problems if the critical failure stress/strain of electrode materials decreases significantly after Li intercalation, while the mechanical design still using the failure criterion for pristine electrode materials. Therefore, it is of critical importance to investigate the mechanical critical failure stress/strain of electrode materials during Li intercalation.

We chose 2D materials black phosphorus (BP) and phosphorene to investigate the Li-concentration dependence of mechanical properties, such as Young's modulus and critical failure stress/strain. BP has a high theoretical Li-storage capacity of 2596 mAhg$^{-1}$, [32] much higher than that of graphene (372 mAhg$^{-1}$) [33]. However, BP suffers from the large volume change of ~300% and rapid capacity loss during charging-discharging cycling as an anode material in LIBs. [34]Phosphorene, the single layer phosphorus, overcoming above disadvantages, is also one optional choice for LIBs due to its reversible capacity of ~433 mAhg$^{-1}$, low open circuit voltage, small volume change, and good electrical conductivity. [35] As a representative 2D electrode material, investigation of the Li-concentration dependence of the mechanical properties of phosphorene could also provide a good reference for other 2D electrode materials.

In this work, we investigated the Li-concentration dependence of elastic modulus and critical failure stress/strain in both phosphorus and phosphorene from the first-principles method. We found the Young's modulus increases with Li intercalations in BP bulk and single layer due to additional formation of Li-P ionic bondings. More importantly, it shows that the critical failure stress/strain will decreases by nearly four times after Li intercalations. This will significantly affect the mechanical bearing capacity and flexibility of encapsulated LIBs during their applications. Our findings suggest that this mechanical deterioration during Li intercalation should be considered carefully in the mechanical design of Li-ion batteries, in order to avoid the mechanical failure in the charge-discharge process.

## Computational Details

All the calculations were performed using Vienna ab initio simulation package (VASP)[36,37], which is based on the density functional theory (DFT). Projector-augmented-wave (PAW) potentials[38] were used taking into account the electron−ion interactions, while the electron exchange-correlation interactions were treated using generalized gradient approximation (GGA)[39] in the scheme of Perdew-Burke-Ernzerhof. A plane wave cutoff of 600 eV was set in our calculations. K-point samplings of $12 \times 8 \times 1$ (monolayer phosphorene) and $12 \times 8 \times 3$ (bulk black phosphorus) were used for calculation. For the monolayer phosphorene, a vacuum space of 20 Å was placed between adjacent layers to avoid mirror interactions. All atomic positions and lattice vectors were fully optimized using a conjugate gradient algorithm to obtain the relaxed configuration. Atomic relaxation was performed until all the forces on each atom are smaller than 0.001 eV/Å. For Li diffusion in phosphorene, we also performed minimum energy path profiling using the climbing image nudged elastic band (CI-NEB) method as implemented in the VASP transition state tools [40,41]. The elastic constant tensor was calculated by the finite difference method in VASP. [42]

## Results and Discussion

**Bulk phosphorus:**

Our calculated lattice constants for bulk black phosphorus are $a$=3.321 Å, $b$=4.421 Å, $c$=10.483 Å, which agrees well with the experimental values ($a$=3.3138 Å, $b$=4.3759 Å, $c$=10.477 Å) [43] and other theoretical calculations ($a$=3.308 Å, $b$=4.536 Å, $c$=11.099 Å) . [44] Fig. 1(a) shows the layered stacking phosphorus structure with Li intercalation. The change of crystal structure for black phosphorus during Li intercalation had been investigated previously. [45] The reactions involved during the first discharge are Black P→$Li_xP$→LiP→$Li_2P$→$Li_3P$. At the initial stage of Li intercalation, pristine phosphorus does not undergo obvious lattice distortion. We chose two intercalation crystal configurations, $x$=0.125 and $x$=0.25 for $Li_xP$, for the further study. Similar to the Li intercalation in graphite, we adopt "stage n" to mark a single

Li-intercalated layer for every $n$ phosphorus sheets, to distinguish different structures for Li intercalated phosphorus structure. For example, stage-2 represents $Li_{0.125}P$ configuration and stage-1 represent $Li_{0.25}P$ configuration, as shown in Fig. 1(b) and 1(c), respectively. Pristine phosphorus is orthorhombic crystal system (space group Cmce), but it reduces to the monoclinic structure after Li intercalation (space group Pm). The lattice parameters of three atomic structure are listed in Table 1. With Li concentration increasing, the phosphorus layer is filled by lithium atoms and the lattice parameter $c$ obviously grows. Meanwhile, lattice parameter $a$ (zigzag direction) shrinks and $b$ (armchair direction) expands slightly with Li atoms intercalated.

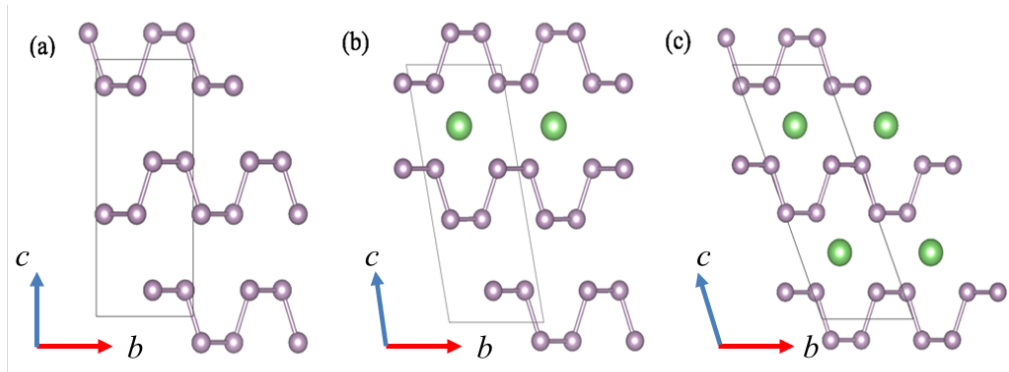

Fig. 1. Atomic structure of Li-intercalated phosphorus bulk (a-c) and single layer (d-f). (a) pristine phosphorus, (b) stage-2 and (c) stage-1 structure. The green and grey balls represent Li and P atoms, respectively. The light grey frame indicates the primitive cell of the structure.

TABLE 1

The lattice parameter of pristine black phosphorus, stage-2 and stage-1 structures.

| Structure | Symmetry group | $a$(Å) | $b$(Å) | $c$(Å) | $\alpha$(°) | $\beta$(°) | $\gamma$(°) |
|---|---|---|---|---|---|---|---|
| Pristine phosphorus | Cmce | 3.32 | 4.42 | 10.48 | 90.00 | 90.00 | 90.00 |
| Stage-2 | Pm | 3.27 | 4.52 | 11.21 | 100.57 | 90.00 | 90.00 |
| Stage-1 | Pm | 3.23 | 4.60 | 12.36 | 111.52 | 90.00 | 90.00 |

We calculated the strain-stress relationship along three directions to investigate the mechanical property of the bulk phosphorus. The strain is defined as $\varepsilon = \frac{a-a_0}{a_0}$, where $a$ and $a_0$ are the lattice constants of the strained and relaxed structure, respectively. The critical stress represents the maximum stress that a structure can withstand. The critical stress can be obtained from the peak of strain-stress curve, while the Young's modulus determines the stiffness of the materials, which is calculated from the elastic constants tensor.

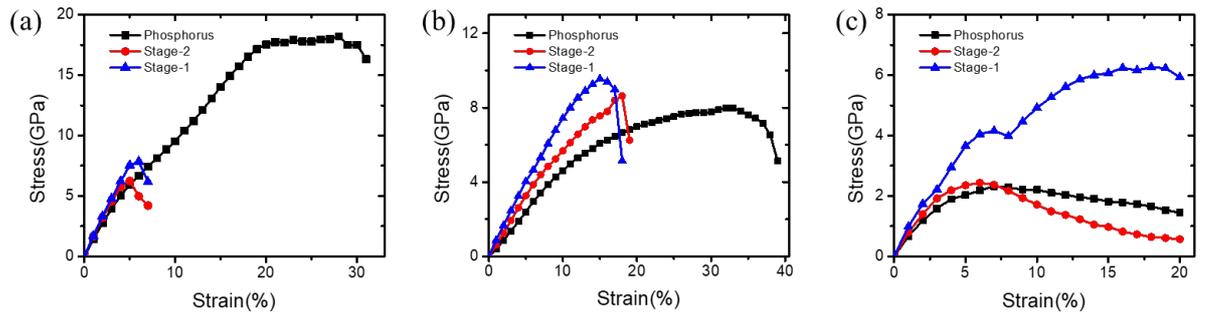

FIG. 2. The strain-stress relations for (a) $x$ direction (zigzag), (b) $y$ direction (armchair) and (c) $z$ direction of pristine phosphorus and Li intercalated structures.

TABLE 2

The Young's module of pristine and Li intercalated phosphorus along zigzag ($x$), armchair ($y$) and out-of-plane ($z$) direction.

| Structure | $E_x$(GPa) | $E_y$(GPa) | $E_z$(GPa) |
| --- | --- | --- | --- |
| Pristine phosphorus | 155.4 | 45.0 | 54.2 |
| Stage-2 | 174.2 | 57.3 | 80.3 |
| Stage-1 | 174.2 | 76.1 | 84.4 |

The strain-stress curves for pristine phosphorus and Li intercalated structures are shown in Fig. 2. It's obvious that the change tendency is anisotropy along three directions. From Fig. 2, we can see that the in-plane critical strain significantly decreases after Li intercalation. For example, compared with the pristine phosphorus the critical strain changes from 28 % to 5 % along zigzag direction and 33 % to 15 % along armchair direction for the configuration with fully filled Li atoms (stage-1). It can also be seen that the critical stress changes from 18.2 GPa to 7.9 GPa along zigzag direction when Li intercalation from zero to fully filled phosphorus atom layer, and 8.0 GPa to 9.6 GPa in armchair direction. The critical stress decreases significantly in zigzag direction and slightly increase in armchair direction. Apparently, the bearing capacity of phosphorus in atom layer plane sharply decrease after Li intercalation. This should be considered carefully during the mechanical design of LIBs because the maximum bearable strain of black phosphorus electrode decreases significantly when charging. We thus cannot adopt the mechanical design criterion which assumes the mechanical properties unchanged for electrode in the whole charge-discharge process.

Figure 2(c) shows that the mechanical properties along z direction (vertical to phosphorus atom plane) is quite different from in-plane properties. For example, the critical stress increases by ~3 times from 2.3 GPa to 6.3 GPa when Li intercalation from zero to fully filled phosphorus atom layer (stage-1), and the corresponding critical strain also significantly increase from 7 % to 18 %. This enhancement can be understood as follows: adjacent phosphorus atom layer combines by weak van der Waals interactions, but it turns to strong ionic bonds between adjacent phosphorus atom layer after introduce intercalation lithium atoms, and therefore, the critical strain/stress increases after charging.

Young's modulus of different Li/P ratio are extracted in Table 2. From Table 2, it can be found that the Young's module monotonically increases with charging. The stiffness along zigzag ($x$) direction increases slightly while it increases tremendously along the other two directions as Li intercalation. For example, the Young's module only increases 12.1% when charging from pristine structure to stage-1 along zigzag direction, but it increases significantly 69.1% and 55.7% along armchair ($y$) and out-of-plane ($z$) direction, respectively. The increase of Young's module indicates Li intercalation strengthens the phosphorus structure in all directions. As more Li intercalation, the number of ionic Li-P bond increases spontaneously. The stronger ionic bonding force replaces the weak van der Waals force in pristine phosphorus and enhance the framework structure. Therefore, the stiffness of black phosphorus increases as Li intercalation.

Interestingly, the change of Young's module in three directions shows strong anisotropy. To explore the reason of these anisotropic changes, we calculated the charge density difference (CDD, $\Delta\rho$) of stage-1, which is shown in Fig. 3. The CDD can be used to appreciate the interaction between the support and the adsorbate. It is defined as:

$$\Delta\rho = \rho_{adsorbate/support} - \rho_{adsorbate} - \rho_{support}$$

where $\rho_{adsorbate/support}$, $\rho_{adsorbate}$ and $\rho_{support}$ are the corresponding charge densities of the combined system ($Li_{0.25}P$), the free adsorbate (Li), and the bare support (P), respectively. It can be seen from Fig.3 that the charge rearrangement after Li intercalation mainly occurs along the armchair and $z$ direction neighboring P atoms of Li, which enhances significantly the bonding interactions along these two directions. Therefore, the Young's module has much larger enhancement in armchair and $z$ directions than zigzag direction. And the charge density projection in $a$ direction is smaller, which means that the effect of ionic bonds is weaker, so the Young's modulus changes less in $a$ direction.

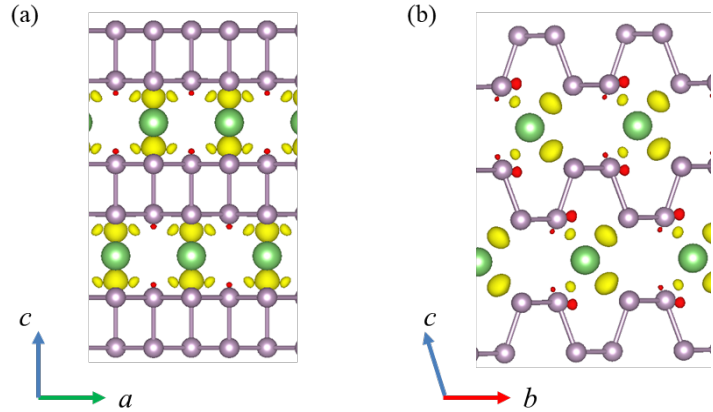

FIG. 3. The charge density difference for stage-1 in two angle of view. The isosurface value is fixed at 0.0035 a.u. with yellow and red colors denoting the charge accumulation and charge depletion, respectively.

**Phosphorene:**

After investigation of the mechanical properties change with Li charging in bulk phosphorus, we then turned to the monolayer phosphorene. The relaxed lattice constants for a monolayer of phosphorene are *a*=3.306Å, *b*=4.574 Å, indicating the shrinkage along *a* and expansion along *b* compared with the bulk phosphorus.

We calculated the mechanical properties of single layer phosphorus, phosphorene. Starting with the relaxed phosphorene structures, tensile strain were applied along either the zigzag or armchair direction to explore its strain-stress relation. Before calculating the mechanical property of Li-intercalated phosphorene, the adsorption position for Li should be determined (Li/P ratio of 0.25, which means fully single-side intercalation). As illustrated in Figure 4(a), the Li atoms were loaded on phosphorene surface. We explored the phosphorene surface for Li binding by placing one Li atom at different sites above phosphorene before structural optimization. After comparing with the total energy in the relaxed structures, it is found that the most stable binding site is above the groove, as shown in Figure 4(a). Specifically, the distance between Li and $P_1$ atom is 2.45 Å, and the distances between Li and $P_2$/$P_3$ atoms are 2.56 Å, which agrees with other simulation results of 2.47 Å and 2.56 Å. [46] For exploring the mechanical properties in different Li/P ratio like bulk structure part, we also relaxed the double-side Li-intercalated phosphorene configuration (the Li/P ratio is 0.5) in the same computing method, which is showed in fig. 4(b). The lattice constant for different Li/P ratio layer structure is listed in Table 3, the lattice constant decrease in zigzag direction and increase in armchair direction as Li/P ratio increasing. The change trend of lattice constant as Li-intercalation for the layer phosphorus is the same as that in bulk.

TABLE 3

The lattice parameter of pristine, single-side Li-intercalated and double-side Li-intercalated phosphorene structures.

| Structure | *a*(Å) | *b*(Å) |
| --- | --- | --- |
| Pristine phosphorene | 3.31 | 4.57 |
| Single-side Li-intercalated | 3.25 | 4.58 |
| Double-side Li-intercalated | 3.24 | 4.74 |

Based on the optimal adsorption site of Li atom, we calculated the strain-stress relation of phosphorene with single- and double-side Li intercalation to explore the variance of mechanical properties for Li intercalation. Since our systems are 2D structures, the elastic constants and moduli from the Hooke's law are under plane-stress conditions [47]

$$\begin{bmatrix} \sigma_{xx} \\ \sigma_{yy} \\ \sigma_{xy} \end{bmatrix} = \frac{1}{1-\nu_{xy}\nu_{yx}} \begin{pmatrix} E_x & \nu_{yx}E_x & 0 \\ \nu_{xy}E_y & E_y & 0 \\ 0 & 0 & G_{xy}(1-\nu_{xy}\nu_{yx}) \end{pmatrix} \begin{bmatrix} \varepsilon_{xx} \\ \varepsilon_{yy} \\ 2\varepsilon_{xy} \end{bmatrix} = \begin{pmatrix} C_{11} & C_{12} & 0 \\ C_{21} & C_{22} & 0 \\ 0 & 0 & C_{66} \end{pmatrix} \begin{bmatrix} \varepsilon_{xx} \\ \varepsilon_{yy} \\ 2\varepsilon_{xy} \end{bmatrix}, \quad (1)$$

Where $\sigma_{ij}$ is the stress tensor, $E_i$ is the Young's modulus in the direction $i$, $C_{ij}$ is elastic constant tensor, $\nu_{ij} = -\dfrac{d\varepsilon_j}{d\varepsilon_i}$ is the Poisson's ratio with strain applied in the direction $i$ and the response strain in the direction $j$, and $G_{xy}$ is the shear modulus in the $xy$ plane. Based on Eq. (1), the relation between the Young's modulus and elastic stiffness constants for a 2D system can be derived

$$E_x = \frac{C_{11}C_{22} - C_{12}C_{21}}{C_{22}}, \quad E_y = \frac{C_{11}C_{22} - C_{12}C_{21}}{C_{11}}. \quad (2)$$

The calculated Young's modulus is 157.8 GPa, 140.1 GPa and 113.8 GPa in the zigzag direction, and 51.5 GPa, 39.3 GPa and 39.2 GPa in the armchair direction for pristine, single-side Li-intercalated and double-side Li-intercalated phosphorene structure, respectively. For the pristine bulk and layer phosphorus, the Young's modulus is approximately equal in both in-plane direction. However, the variation of the Young's modulus as Li-intercalation is opposite for bulk and layer phosphorus, the Young's modulus increases as Li-intercalation in both in-plane direction for bulk phosphorus while it decreases for layered phosphorus. This is because there is no van der Waals effect in single layered phosphorus and the ionic bonds introduced by intercalated lithium atoms weaken the covalent bond between phosphorus atoms (it is further discussed below), which reduces the modulus.

Our calculated strain-stress relations of pristine, single-side Li-intercalated (Li/P ratio of 0.25) and double-side Li-intercalated (Li/P ratio of 0.5) phosphorene are presented in Figure 4(c) and (d). We also summarized the critical stress and strain for both bulk and single layer phosphorus in Table 4. For the pristine phosphorene, the critical stress (strain) is 23.2 GPa (25%) and 13.4 GPa (33%) along the zigzag and armchair directions, respectively. Compared with pristine bulk phosphorus, it can be seen that the critical stress of single layered pristine phosphorus increases significantly from 18.2 GPa to 23.2 GPa (zigzag direction) and 7.9 GPa to 13.4 GPa (armchair direction), while the critical strain nearly keeps the same.

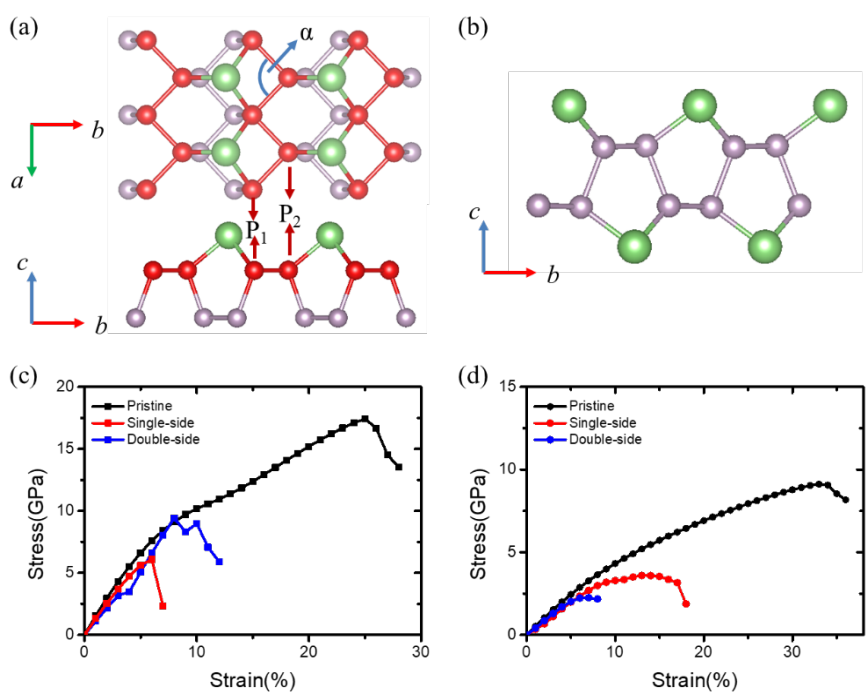

FIG. 4. (a) The configure of single-side Li-intercalated phosphorene. The green and grey balls represent Li and P atoms, respectively, and the top layer of P atoms is shown as red balls. (b) The configure of double-side Li-intercalated phosphorene. The strain-stress relation for phosphorene and Li-intercalated phosphorene along zigzag (c) and armchair (d) directions.

After the lithium intercalation in the single-side of phosphorene, compared with pristine phosphorene, the critical stress remarkably decreases from 23.2 GPa to 2.6 GPa (nearly 9 times) and 13.4 GPa to 3.7 GPa (more than 3 times) in the zigzag and armchair directions, respectively. The critical strain also has significantly decreases, i.e., from

25% to 6% in zigzag direction and from 33% to 14% in armchair direction. As can be seen, similar to the bulk phosphorus, the lithium intercalation can seriously weaken the mechanical bearing capacity of layered phosphorus. This mechanical weakening effect induced by Li charging should be taken into account carefully during the mechanical design of LIBs, in order to keep LIBs work safely during the charge-discharge process.

TABLE 4

The ideal strength and critical strain in various strain direction for BP and phosphorene w/o Li intercalation, respectively. The "Bulk" in this table means pristine BP, and the "Li-Bulk", "Layer", "Li-Layer" means stage-1 for BP, phosphorene, single-side Li-intercalated phosphorene, respectively.

|  | **Ideal Strength (GPa)** | | **Critical strain (%)** | |
| :---: | :---: | :---: | :---: | :---: |
|  | zigzag | armchair | zigzag | armchair |
| Bulk | 18.2 | 8.0 | 28 | 33 |
| Li-Bulk | 7.9 | 9.6 | 5 | 15 |
| Layer | 23.2 | 13.4 | 25 | 33 |
| Li-Layer | 2.6 | 3.7 | 6 | 14 |

The remarkable critical strain/stress reduction may be explained by the change of electronic properties for lithium intercalated. Phosphorus has $3s^2 3p^3$ valence electron configuration, forms $sp^3$ bonding with a lone pair of valence electrons in each phosphorus atom, which does not participate in the direct bonding with other atoms. [48] After Li-intercalation, the intercalated lithium atoms influence the pristine electronic properties and change the stability of structure. We performed a projected crystal orbital Hamiltonian population (pCOHP) bonding analysis[49-51] to examine the difference in bonding properties between pristine and Li-intercalated phosphorene. The pCOHP decomposes the density of state (DOS) according to the weighted Hamiltonian matrix elements. Bonding and antibonding states are represented by positive and negative values of –pCOHP, respectively. We choose the labeled phosphorus atoms $P_1$ and $P_2$ in figure 5(a) to calculate the pCOHP, because they form ionic bonds with the intercalated lithium atoms. The figure 5(a) and 5(b) show the COHP of bond $P_1$-$P_2$ for pristine and single-side Li-intercalated phosphorene. A finite antibonding interaction is found at the Fermi level (Fig. 5b), indicating that the covalent effect weakens. In addition, the integrated COHP (ICOHP) can provide an estimation of the strength of bonding. The ICOHP of bond $P_1$-$P_2$ is -4.98 and -4.10 eV for pristine and single-side Li-intercalated phosphorene, suggesting the covalent bond become weaker after Li-intercalation. We also calculated the partial density of states (PDOS) of pristine and single-side Li-intercalated configure, as shown Fig. 5(c) and (d). It can be seen that the DOS decreases as Li intercalation, especially the overlap area of *s* and *p* orbitals reduces, suggesting the $sp^3$ hybrid bonds become weak after lithium atoms intercalation. Therefore, the intercalated lithium atoms change the electronic properties of pristine phosphorene and weaken the $sp^3$ hybrid bonds, which causes the remarkable mechanical properties deterioration.

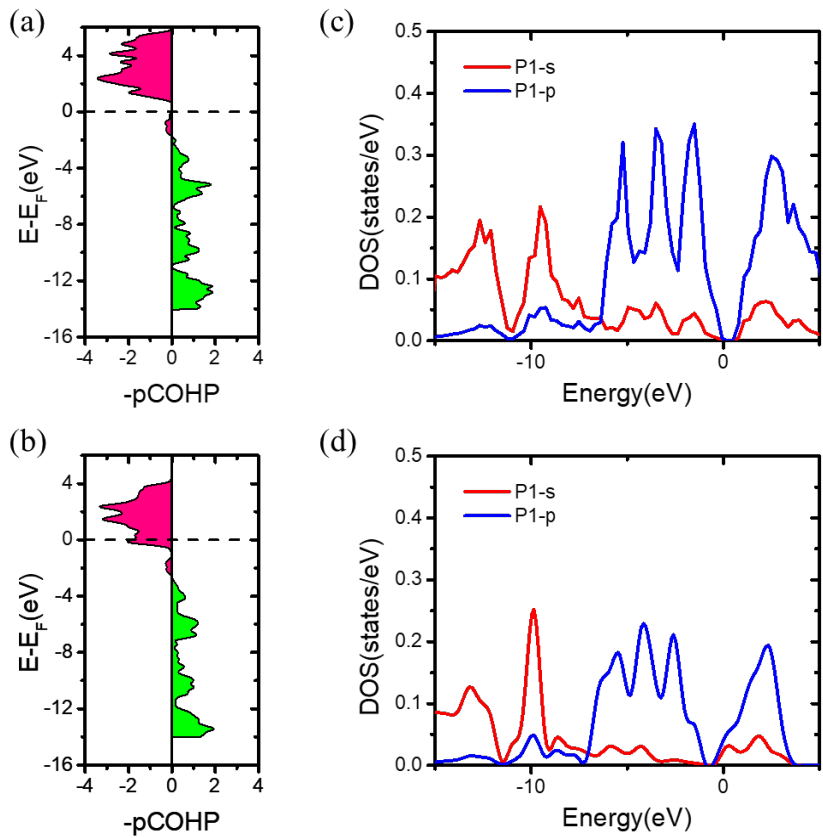

FIG. 5. The COHP of bond $P_1$-$P_2$ for (a) pristine and (b) single-side Li-intercalated phosphorene, the energy axis is shown relative to the Fermi level ($E_F$). The PDOS for

P$_1$ in (c) pristine and (d) single-side Li-intercalated phosphorene.

**Lithium diffusions in phosphorene under strains:**

We showed that Li intercalation can significantly affect the mechanical properties of bulk and single layer phosphorus. Next we will investigate how the external mechanical loadings impact on lithium diffusion and migration. As is known, the diffusion of lithium on phosphorene is highly anisotropic with diffusion along the zigzag direction being highly energetically favorable and along the armchair direction being almost prohibited, due to the phosphorene puckered structure. [46] We firstly calculated lithium migration energy profile along the zigzag direction in pristine phosphorene without strain, as it's shown in Figure 6(a). The energy barrier of phosphorene is 118 meV which agrees well with the previous work. [46]

To investigate the effect of strain on the dynamics of lithium atoms in phosphorene, we calculated the variation of the energy barriers as a function of the external strain along zigzag and armchair directions, which is shown in Fig. 6(b). As can be seen, the energy barrier decreases (increases) with compressive (tensile) strain along zigzag direction and/or tensile (compressive) strain along armchair direction in the strain range of -4% ~ 4%. Therefore, it will be easier for lithium diffusion and improve the Li ion conductivity of phosphorene by applying proper compressive strain in zigzag direction and/or tensile strain in armchair direction. This can be easily understood: tensile strain along armchair direction (or compressive strain along zigzag direction, due to Poisson's effect) widens the groove and therefore weakens the interactions between P and Li atoms, which makes it easier for Li ions transport along the groove (zigzag direction). This suggests that one could tune the ion transport properties of phosphorene electrode by applying proper mechanical loadings.

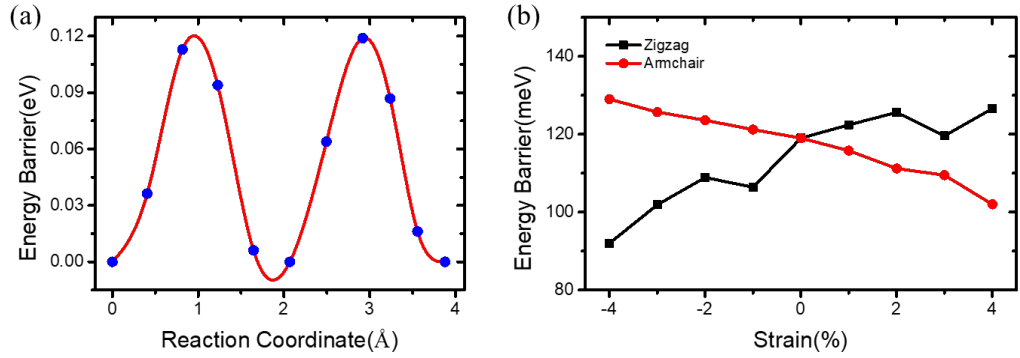

FIG. 6. (a) Energy profiles of Li diffusion in pristine phosphorene without strain. (b) The variation of the energy barriers as a function of the applied external strain along zigzag and armchair directions.

Furthermore, similar to previous work, [52] we also found that the ferroelastic phase transition occurs, i.e., the zigzag and armchair directions switch, when applying more than 30% strain in zigzag direction and then relaxing the structure. As Li ions transfer in phosphorene surface in zigzag direction with armchair direction having much higher energy barrier, it would expect that the Li ion transfer path will also has 90-degree rotation after this ferroelastic phase transition under strain. This may be applied to tune the Li ion transfer path of the LIBs by external strain. However, there will be a long way to utilize this interesting effect because it is challenging to apply such large strain in the real LIBs applications.

## Conclusion

In conclusion, we have computed the mechanical properties of both bulk and single layer phosphorus during the lithium intercalation process, and the Li ion diffusion energy barrier for phosphorene with different applied strain by using the first-principles calculations. Our results show that the Young's modulus of bulk phosphorus are improved as lithium intercalation due to increase of Li-P ionic bonds in the systems. And the effect is most obvious in the vertical to phosphorus atom layer and armchair direction, in which direction the Young's modulus becomes nearly two times. However, the mechanical bearing capacities, such as critical strain and stress, are significantly reduced by several times after lithium intercalation in both bulk and single layer phosphorus, which may reduce the reliability of Li-ion batteries. The Li intercalation introducing ionic bonds to phosphorus causes the significant mechanical property change. Our findings suggest that the mechanical properties are not invariant in charge-discharge process, this remarkable mechanical properties deterioration during Li intercalation should be considered in the mechanical design of Li-ion batteries. Besides, strain in zigzag direction increases the Li ion diffusion energy barrier, while in armchair direction deceases, we expect that this strain-engineering can be a promising approach to optimize lithium-based energy storage for electric vehicles requiring high power. Furthermore, bulk phosphorus and layered phosphorene are typical electrode materials with interlayer van der Waals effect. We would expect that the lithium intercalation could also induce the mechanical properties deterioration in other layered electrode materials.

## Acknowledgements

The research support of the National Materials Genome Project (2016YFB0700600), the Foundation for Innovative Research Groups of the National Natural Science Foundation of China (No. 11521202, No. 11572040 and No. 11804023), the China Postdoctoral Science Foundation with Grant No. 2018M641205, the Thousand Young Talents Program of China, National Key Research and Development Program of China (Grant No. 2016YFB0402700), and Graduate Technological Innovation Project of Beijing Institute of Technology is gratefully acknowledged.

## Reference

[1]. Whittingham M S. Electrical energy storage and intercalation chemistry[J]. Science, 1976, 192(4244): 1126-1127.

[2]. Besenhard J O, Eichinger G. High energy density lithium cells: Part I. Electrolytes and anodes[J]. Journal of Electroanalytical Chemistry and Interfacial Electrochemistry,

1976, 68(1): 1-18.

[3]. Eichinger G, Besenhard J O. High energy density lithium cells: Part II. Cathodes and complete cells[J]. Journal of Electroanalytical Chemistry and Interfacial Electrochemistry, 1976, 72(1): 1-31.

[4]. Simon P, Gogotsi Y. Materials for electrochemical capacitors[M]//Nanoscience And Technology: A Collection of Reviews from Nature Journals. 2010: 320-329.

[5]. Arico A S, Bruce P, Scrosati B, et al. Nanostructured materials for advanced energy conversion and storage devices[M]//Materials for sustainable energy: a collection of peer-reviewed research and review articles from Nature Publishing Group. 2011: 148-159.

[6]. Whittingham M S. Lithium batteries and cathode materials[J]. Chemical reviews, 2004, 104(10): 4271-4302.

[7]. Liu C, Li F, Ma L P, et al. Advanced materials for energy storage[J]. Advanced materials, 2010, 22(8): E28-E62.

[8]. El-Kady M F, Strong V, Dubin S, et al. Laser scribing of high-performance and flexible graphene-based electrochemical capacitors[J]. Science, 2012, 335(6074): 1326-1330.

[9]. Tamura N, Fujimoto M, Kamino M, et al. Mechanical stability of Sn–Co alloy anodes for lithium secondary batteries[J]. Electrochimica Acta, 2004, 49(12): 1949-1956.

[10]. Li K, Xie H, Liu J, et al. From chemistry to mechanics: bulk modulus evolution of Li–Si and Li–Sn alloys via the metallic electronegativity scale[J]. Physical Chemistry Chemical Physics, 2013, 15(40): 17658-17663.

[11]. An Y, Wood B C, Ye J, et al. Mitigating mechanical failure of crystalline silicon electrodes for lithium batteries by morphological design[J]. Physical Chemistry Chemical Physics, 2015, 17(27): 17718-17728.

[12]. Beaulieu L Y, Eberman K W, Turner R L, et al. Colossal reversible volume changes in lithium alloys[J]. Electrochemical and Solid-State Letters, 2001, 4(9): A137-A140.

[13]. Liu N, Lu Z, Zhao J, et al. A pomegranate-inspired nanoscale design for large-

volume-change lithium battery anodes[J]. Nature nanotechnology, 2014, 9(3): 187.

[14]. Beaulieu L Y, Hatchard T D, Bonakdarpour A, et al. Reaction of Li with alloy thin films studied by in situ AFM[J]. Journal of The Electrochemical Society, 2003, 150(11): A1457-A1464.

[15]. Grugeon S, Laruelle S, Herrera-Urbina R, et al. Particle size effects on the electrochemical performance of copper oxides toward lithium[J]. Journal of The Electrochemical Society, 2001, 148(4): A285-A292.

[16]. Cheng Y T, Verbrugge M W. Diffusion-induced stress, interfacial charge transfer, and criteria for avoiding crack initiation of electrode particles[J]. Journal of the Electrochemical Society, 2010, 157(4): A508-A516.

[17]. Bucci G, Swamy T, Chiang Y M, et al. Modeling of internal mechanical failure of all-solid-state batteries during electrochemical cycling, and implications for battery design[J]. Journal of Materials Chemistry A, 2017, 5(36): 19422-19430.

[18]. Park S, Vosguerichian M, Bao Z. A review of fabrication and applications of carbon nanotube film-based flexible electronics[J]. Nanoscale, 2013, 5(5): 1727-1752.

[19]. Deshpande R, Cheng Y T, Verbrugge M W. Modeling diffusion-induced stress in nanowire electrode structures[J]. Journal of Power Sources, 2010, 195(15): 5081-5088.

[20]. Endo M, Kim Y A, Hayashi T, et al. Vapor-grown carbon fibers (VGCFs): basic properties and their battery applications[J]. Carbon, 2001, 39(9): 1287-1297.

[21]. Cheng Y T, Verbrugge M W. Evolution of stress within a spherical insertion electrode particle under potentiostatic and galvanostatic operation[J]. Journal of Power Sources, 2009, 190(2): 453-460.

[22]. Christensen J, Newman J. A mathematical model of stress generation and fracture in lithium manganese oxide[J]. Journal of The Electrochemical Society, 2006, 153(6): A1019-A1030.

[23]. Zhang X, Sastry A M, Shyy W. Intercalation-induced stress and heat generation within single lithium-ion battery cathode particles[J]. Journal of The Electrochemical Society, 2008, 155(7): A542-A552.

[24]. Cheng Y T, Verbrugge M W. The influence of surface mechanics on diffusion induced stresses within spherical nanoparticles[J]. Journal of Applied Physics, 2008,

104(8): 083521.

[25]. Shenoy V B, Johari P, Qi Y. Elastic softening of amorphous and crystalline Li–Si phases with increasing Li concentration: a first-principles study[J]. Journal of Power Sources, 2010, 195(19): 6825-6830.

[26]. Qi Y, Guo H, Hector L G, et al. Threefold increase in the Young's modulus of graphite negative electrode during lithium intercalation[J]. Journal of The Electrochemical Society, 2010, 157(5): A558-A566.

[27]. Yang F Q. Diffusion-induced stress in inhomogeneous materials: concentration-dependent elastic modulus[J]. Science China Physics, Mechanics and Astronomy, 2012, 55(6): 955-962.

[28]. Mortazavi M, Deng J, Shenoy V B, et al. Elastic softening of alloy negative electrodes for Na-ion batteries[J]. Journal of Power Sources, 2013, 225: 207-214.

[29]. Zhang J, Lu B, Song Y, et al. Diffusion induced stress in layered Li-ion battery electrode plates[J]. Journal of Power Sources, 2012, 209: 220-227.

[30]. Li Y, Yang J, Song J. Nano-energy system coupling model and failure characterization of lithium ion battery electrode in electric energy vehicles[J]. Renewable and Sustainable Energy Reviews, 2016, 54: 1250-1261.

[31]. Kalnaus S, Rhodes K, Daniel C. A study of lithium ion intercalation induced fracture of silicon particles used as anode material in Li-ion battery[J]. Journal of Power Sources, 2011, 196(19): 8116-8124.

[32]. Nagao M, Hayashi A, Tatsumisago M. All-solid-state lithium secondary batteries with high capacity using black phosphorus negative electrode[J]. Journal of Power Sources, 2011, 196(16): 6902-6905.

[33]. Tarascon J M, Armand M. Issues and challenges facing rechargeable lithium batteries[M]//Materials for Sustainable Energy: A Collection of Peer-Reviewed Research and Review Articles from Nature Publishing Group. 2011: 171-179.

[34]. Batmunkh M, Bat- Erdene M, Shapter J G. Phosphorene and Phosphorene-Based Materials–Prospects for Future Applications[J]. Advanced Materials, 2016, 28(39): 8586-8617.

[35]. Zhang C, Yu M, Anderson G, et al. The prospects of phosphorene as an anode

material for high-performance lithium-ion batteries: a fundamental study[J]. Nanotechnology, 2017, 28(7): 075401.

[36]. Kresse G, Furthmüller J. Efficient iterative schemes for ab initio total-energy calculations using a plane-wave basis set[J]. Physical review B, 1996, 54(16): 11169.

[37]. Kresse G, Furthmüller J. Efficiency of ab-initio total energy calculations for metals and semiconductors using a plane-wave basis set[J]. Computational materials science, 1996, 6(1): 15-50.

[38]. Blöchl P E. Projector augmented-wave method[J]. Physical review B, 1994, 50(24): 17953.

[39]. Perdew J P, Chevary J A, Vosko S H, et al. Atoms, molecules, solids, and surfaces: Applications of the generalized gradient approximation for exchange and correlation[J]. Physical review B, 1992, 46(11): 6671.

[40]. Henkelman G, Uberuaga B P, Jónsson H. A climbing image nudged elastic band method for finding saddle points and minimum energy paths[J]. The Journal of chemical physics, 2000, 113(22): 9901-9904.

[41]. Henkelman G, Jónsson H. Improved tangent estimate in the nudged elastic band method for finding minimum energy paths and saddle points[J]. The Journal of chemical physics, 2000, 113(22): 9978-9985.

[42]. Le Page Y, Saxe P. Symmetry-general least-squares extraction of elastic data for strained materials from ab initio calculations of stress[J]. Physical Review B, 2002, 65(10): 104104.

[43]. Brown A, Rundqvist S. Refinement of the crystal structure of black phosphorus[J]. Acta Crystallographica, 1965, 19(4): 684-685.

[44]. Wei Q, Peng X. Superior mechanical flexibility of phosphorene and few-layer black phosphorus[J]. Applied Physics Letters, 2014, 104(25): 251915.

[45]. Park C M, Sohn H J. Black phosphorus and its composite for lithium rechargeable batteries[J]. Advanced materials, 2007, 19(18): 2465-2468.

[46]. Li W, Yang Y, Zhang G, et al. Ultrafast and directional diffusion of lithium in phosphorene for high-performance lithium-ion battery[J]. Nano letters, 2015, 15(3): 1691-1697.

[47]. Zhou J, Huang R. Internal lattice relaxation of single-layer graphene under in-plane deformation[J]. Journal of the Mechanics and Physics of Solids, 2008, 56(4): 1609-1623.

[48]. Dai J, Zeng X C. Structure and stability of two dimensional phosphorene with [double bond, length as m-dash] O or [double bond, length as m-dash] NH functionalization[J]. Rsc Advances, 2014, 4(89): 48017-48021.

[49]. Dronskowski R, Blöchl P E. Crystal orbital Hamilton populations (COHP): energy-resolved visualization of chemical bonding in solids based on density-functional calculations[J]. The Journal of Physical Chemistry, 1993, 97(33): 8617-8624.

[50]. Grechnev A, Ahuja R, Eriksson O. Balanced crystal orbital overlap population—a tool for analysing chemical bonds in solids[J]. Journal of Physics: Condensed Matter, 2003, 15(45): 7751.

[51]. Steinberg S, Dronskowski R. The crystal orbital Hamilton population (COHP) method as a tool to visualize and analyze chemical bonding in intermetallic compounds[J]. Crystals, 2018, 8(5): 225.

[52]. Wu M, Zeng X C. Intrinsic ferroelasticity and/or multiferroicity in two-dimensional phosphorene and phosphorene analogues[J]. Nano letters, 2016, 16(5): 3236-3241.